\documentclass[]{fairmeta}

\title{How Well Does Generative Recommendation Generalize?}

\author[1]{Yijie Ding}
\author[2]{Zitian Guo}
\author[3]{Jiacheng Li}
\author[2]{Letian Peng}
\author[3]{Shuai Shao}
\author[3]{Wei Shao}
\author[3]{Xiaoqiang Luo}
\author[3]{Luke~Simon}
\author[2]{Jingbo Shang}
\author[2]{Julian McAuley}
\author[2]{Yupeng Hou}

\affiliation[1]{Carnegie Mellon University}
\affiliation[2]{University of California San Diego}
\affiliation[3]{Meta}

\date{\today}
\correspondence{Yupeng Hou at \email{yphou@ucsd.edu}}

\metadata[Code]{\url{https://github.com/Jamesding000/MemGen-GR}}

\usepackage{xspace}
\usepackage{booktabs}
\usepackage{multirow}
\usepackage{graphicx}
\usepackage{xcolor}
\usepackage{tcolorbox}
\usepackage{colortbl}
\usepackage{amsmath}
\usepackage{subcaption}
\usepackage{enumitem}
\usepackage{xstring}
\usepackage{etoolbox}
\usepackage{amssymb}
\usepackage{adjustbox}
\usepackage{wrapfig}

\definecolor{GrayBlue}{HTML}{EBF5FF}
\definecolor{Gray}{gray}{0.9}
\definecolor{PerfRed}{RGB}{192, 57, 43}
\definecolor{PerfGreen}{RGB}{39, 174, 96}

\newcommand{\ie}{\emph{i.e.,}\xspace}
\newcommand{\eg}{\emph{e.g.,}\xspace}

\newcommand{\ignore}[1]{}
\newcommand{\paratitle}[1]{\vspace{1.5ex}\noindent\textbf{#1}}

\makeatletter
\patchcmd{\@mkbibcitation}
  {\authors\egroup. \@acmYear. \@title
   \ifx\@subtitle\@empty. \else: \@subtitle. \fi}
  {\authors\egroup. \@acmYear. \@title
   \IfEndWith{\@title}{?}
     {\ifx\@subtitle\@empty\ \else: \@subtitle. \fi}
     {\ifx\@subtitle\@empty.\else: \@subtitle. \fi}}
  {}{}
\makeatother

\abstract{
A widely held hypothesis for why generative recommendation (GR) models outperform conventional item ID-based models is that they generalize better. However, there is few systematic way to verify this hypothesis beyond a superficial comparison of overall performance. To address this gap, we categorize each data instance based on the specific capability required for a correct prediction: either memorization (reusing item transition patterns observed during training) or generalization (composing known patterns to predict unseen item transitions). Extensive experiments show that GR models perform better on instances that require generalization, whereas item ID-based models perform better when memorization is more important. To explain this divergence, we shift the analysis from the item level to the token level and show that what appears to be item-level generalization often reduces to token-level memorization for GR models. Finally, we show that the two paradigms are complementary. We propose a simple memorization-aware indicator that adaptively combines them on a per-instance basis, leading to improved overall recommendation performance.
}

\begin{document}

\maketitle

\begingroup
\renewcommand\thefootnote{}
\footnotetext{The experiments in this work were conducted using computing resources provided by UC San Diego.}
\addtocounter{footnote}{-1}
\endgroup

\section{Introduction}
\label{sec:intro}


\begin{figure*}[t]
    \centering
    \includegraphics[width=0.98\textwidth]{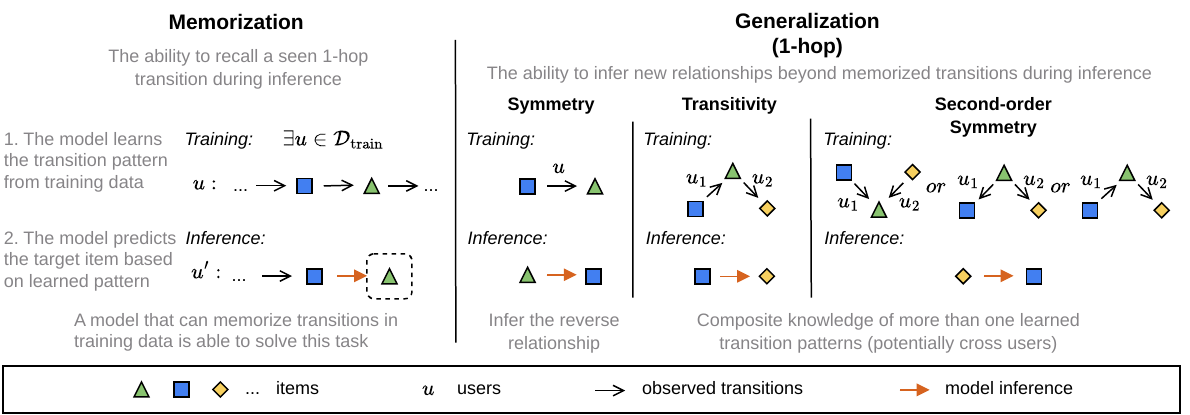}
    \caption{Illustrated definitions for memorization vs. generalization. We define memorization and different sub-categories of generalization based on (1) the transition patterns observed in training data, and (2) the patterns required to infer.
    }
    \label{fig:definitions}
\end{figure*}

Generative recommendation (GR)~\citep{rajput2023tiger,zheng2024lcrec,deng2025onerec,he2025plum} has recently emerged as a promising paradigm for sequential recommendation. Compared with conventional models such as SASRec~\citep{kang2018sasrec}, a key difference is that GR models tokenize each item as a sequence of sub-item tokens (\eg semantic IDs~\citep{tay2022dsi,rajput2023tiger}) rather than a single unique item ID.
However, the advantage of GR models has typically been observed in terms of overall performance, that is, GR models correctly predict more data instances than conventional methods~\citep{rajput2023tiger,deng2025onerec}. This naturally raises the question of which types of data instances are better handled by generative recommendation models.

We hypothesize that each data instance requires different levels of generalization and memorization for correct prediction, leading to the performance discrepancies observed between GR and item ID-based models.
To investigate this, we propose an analytical framework that categorizes each data instance by the primary model capability it requires (either memorization or generalization) based on the underlying data patterns.
We then analyze model performance on each category separately. 
Nevertheless, conducting such analyses requires two key components: identifying the data patterns of interest in the context of sequential recommendation, and designing reasonable methods to categorize instances.

\textbf{(1) Data patterns.} Since the task is framed as predicting the next item from a user's history, a natural starting point is to focus on the target items. While prior work often studies cold-start items (\ie items that are rare or unseen during training) as out-of-distribution cases that require generalization~\citep{singh2024spmsid,yang2025liger,ding2026specgr}, this target-centric view ignores the interaction between the history and the target. Even when a target item is popular, the transition from the given history to that item may be rare in the training data. Predicting such transitions can therefore still require generalization.

\textbf{(2) Categorization.} 
We require a principled method to determine whether a given instance primarily relies on memorization or generalization.
Prior studies, such as those based on counterfactual memorization~\citep{zhang2023counterfactual,grosse2023studying,raunak2021curious,ghosh2025rethinking}, are usually computationally expensive, as they require frequent model retraining on datasets that exclude specific data points.
This makes them impractical for recommendation settings with large-scale user interaction logs~\citep{deng2025onerec,zhai2024hstu}. Another line of work categorizes instances by
measuring representation similarities between training instances and the predictions~\citep{ivison2025large,pezeshkpour2021empirical,pruthi2020estimating}. However, these methods are mainly adopted in tasks without a clear ground truth, such as language modeling. In contrast, recommendation is typically evaluated with a well-defined ground-truth target item for each instance~\citep{kang2018sasrec,he2017neural}.

Given these considerations, we treat \emph{item transitions} (from a historical item to the target) rather than single target items as the data patterns of interest  (\Cref{fig:definitions}).
To categorize data instances, we examine whether the item transitions required for the correct prediction have been observed in the training data (memorization), or if they can be composed or inferred from observed patterns (generalization). 
Using this categorization, we explicitly partition the test data into subsets reflecting different capabilities and evaluate model performance on each, thereby distinguishing the contributions of memorization and generalization to overall performance.

To this end, we benchmark two representative models for each paradigm: TIGER~\citep{rajput2023tiger} as the semantic ID-based GR model and SASRec~\citep{kang2018sasrec} as the item ID-based conventional model. By evaluating performance on memorization and generalization subsets across seven real-world datasets, we find that GR models indeed excel on generalization-related subsets, while generally underperforming item ID-based models on memorization-related subsets. This observation motivates us to investigate the mechanism behind the generalization capability of GR models. 
We then shift our analysis from item transition patterns to sub-item token transition patterns. From this perspective, a substantial fraction of target item transitions that would be regarded as item-level generalization can instead be interpreted as token-level memorization. 
This effectively explains the source of the GR models' generalization capability.

Finally, we show that these two paradigms are complementary. We introduce an adaptive ensembling method that combines a GR model with an item ID-based model. The ensemble assigns instance-specific weights based on whether each data instance primarily requires memorization or generalization, as predicted by an indicator. Experimental results show that this adaptive ensembling strategy consistently improves overall performance over both individual models and naive fixed-weight ensembles.


\section{Defining Memorization and Generalization}
\label{sec:defining-memorization-and-generalization}

In this section, we describe our proposed framework for analyzing memorization and generalization in sequential recommendation. We first outline the task definition and notation in~\Cref{subsec:task}. In what follows, we present our criteria for attributing data instances as memorization-based (\Cref{subsec:mem}) or generalization-based (\Cref{subsec:gen}), relying on the item transition patterns they contain. Subsequently, we extend the generalization criteria to encompass multi-hop generalization in~\Cref{subsec:multi-hop-gen}. Finally, we discuss the remaining uncategorized instances in~\Cref{subsec:uncategorized}.

\subsection{Task Definition}\label{subsec:task}


\paratitle{Sequential recommendation.} A user is represented by a sequence of historical item interactions $u = [i_1, i_2, \ldots, i_{t-1}]$, where $i \in \mathcal{I}$. The goal is to predict the next item $i_t$ that the user will interact with. The recommendation models are trained on a set of user interaction sequences $\mathcal{D}_{\text{train}}$. For a data instance $(u, i_t)$ not present in the training set, we attribute it to memorization or generalization based on its constituent item transition patterns.

\paratitle{Item transition.} As discussed in~\Cref{sec:intro}, we treat item transitions as the fundamental data patterns for studying memorization and generalization. Specifically, we define an item transition as a directed pair of items $[i_s \to i_t]$, where $i_s, i_t \in u$ and $s < t$. We further define the \emph{hop count} of the item transition based on the distance between $i_s$ and $i_t$ in the user's history. For example, if $s = t-2$, the hop count is $2$.
Our framework categorizes each data instance $(u, i_t)$ based on the set of item transitions $\{[i_s \to i_t], i_s \in u \}$ it contains.

\subsection{Memorization-Related Data Instance}\label{subsec:mem}

We define a data instance as \emph{memorization-related} if the $1$-hop item transition $[i_{t-1} \to i_t]$ has been observed in the training data, regardless of which user's history it appears in. Under this condition, it's possible for a model to correctly predict the target item only by memorizing the training data.

\begin{tcolorbox}[colback=GrayBlue,colframe=GrayBlue,arc=5pt,boxrule=0pt,left=2pt,right=2pt,top=0pt,bottom=0pt]
\textbf{Def. (Memorization).} A data instance $(u, i_t)$ is \textit{memorization-related} if:
\begin{equation*}
    (u, i_t) \in \mathcal{D}_{\text{mem}} \iff \exists u' \in \mathcal{D}_{\text{train}} \text{ s.t. } [i_{t-1} \to i_t] \subseteq u'.
\end{equation*}
\end{tcolorbox}


\subsection{Generalization-Related Data Instance}\label{subsec:gen}

A data instance is defined as \emph{generalization-related} if: (1) it is not memorization-related; and (2) it contains at least one item transition that can be inferred or composed from observed transitions in the training data. We categorize generalization into multiple types based on specific inference or composition methods. Note that a single data instance may satisfy multiple generalization types.

For simplicity, we first focus on $1$-hop item transitions $[i_s \to i_t]$ where $s = t-1$, introducing three possible $1$-hop generalization types: transitivity, symmetry, and 2nd-order symmetry.

\paratitle{Transitivity} implies that the model can infer $[i_{t-1} \to i_t]$ by bridging two observed transitions via an intermediate item $x$.

\begin{tcolorbox}[colback=GrayBlue,colframe=GrayBlue,arc=5pt,boxrule=0pt,left=2pt,right=2pt,top=0pt,bottom=0pt]
\textbf{Def. (Transitivity).} A data instance $(u, i_t) \notin \mathcal{D}_{\text{mem}}$ satisfies \textit{transitivity} if:
\begin{equation*}
    \exists x \in \mathcal{I}, \exists u', u'' \in \mathcal{D}_{\text{train}}, [i_{t-1} \to x] \subseteq u' \land [x \to i_t] \subseteq u''.
\end{equation*}
\end{tcolorbox}

\paratitle{Symmetry} allows the model to infer a transition $[i_{t-1} \to i_t]$ if its reverse $[i_t \to i_{t-1}]$ has been observed.

\begin{tcolorbox}[colback=GrayBlue,colframe=GrayBlue,arc=5pt,boxrule=0pt,left=2pt,right=2pt,top=0pt,bottom=0pt]
\textbf{Def. (Symmetry).} A data instance $(u, i_t) \notin \mathcal{D}_{\text{mem}}$ satisfies \textit{symmetry} if:
\begin{equation*}
    \exists u' \in \mathcal{D}_{\text{train}}, [i_t \to i_{t-1}] \subseteq u'.
\end{equation*}
\end{tcolorbox}

\paratitle{2nd-Order Symmetry} encompasses complex symmetric relations where $i_{t-1}$ and $i_t$ are related via an intermediate item $x$ in non-transitive ways.

\begin{tcolorbox}[colback=GrayBlue,colframe=GrayBlue,arc=5pt,boxrule=0pt,left=2pt,right=2pt,top=0pt,bottom=0pt]
\textbf{Def. (2nd-Order Symmetry).} A data instance $(u, i_t) \notin \mathcal{D}_{\text{mem}}$ satisfies \textit{2nd-order symmetry} if $\exists x \in \mathcal{I}, \exists u', u'' \in \mathcal{D}_{\text{train}}$ such that one of the following holds:
\begin{align*}
    (1) & \ [x \to i_{t-1}] \subseteq u' \land [x \to i_t] \subseteq u''  & (\text{common cause}); \\
    (2) & \ [i_{t-1} \to x] \subseteq u' \land [i_t \to x] \subseteq u''  & (\text{common effect}); \\
    (3) & \ [i_t \to x] \subseteq u' \land [x \to i_{t-1}] \subseteq u''  & (\text{reverse path}).
\end{align*}
\end{tcolorbox}

\subsection{Multi-Hop Generalization}\label{subsec:multi-hop-gen}

\begin{wrapfigure}{r}{0.52\linewidth}
    \centering
    \vspace{-20pt}
    \includegraphics[width=1.0\linewidth]{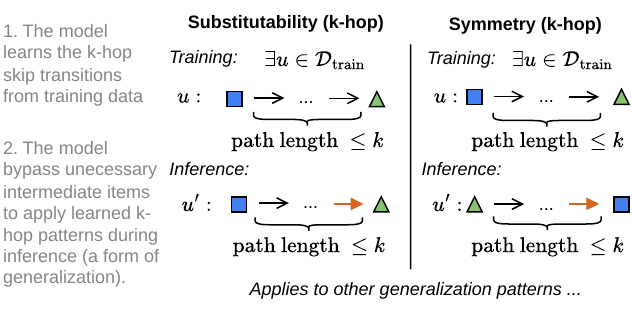}
    \caption{Illustration of multi-hop generalization.
    }
    \label{fig:multi-hop-extension}
    \vspace{-20pt}
\end{wrapfigure}

Although \Cref{subsec:gen} focuses on $1$-hop item transitions for simplicity, our proposed criteria naturally extend to multi-hop transitions. Specifically, we define multi-hop generalization types (transitivity, symmetry, and 2nd-order symmetry) by applying the same logic to multi-hop item transitions (as defined in~\Cref{subsec:task}). If a data instance involves multiple item transitions with different hop counts that satisfy the generalization criteria, we use the minimum hop count for categorization.

\paratitle{Substitutability.} Beyond the types introduced in~\Cref{subsec:gen}, one might consider extending the definition of memorization to multi-hop item transitions. However, we argue that ``multi-hop memorization'' is effectively a form of generalization. It requires the model to have strong generalization capabilities to bypass unnecessary intermediate items and select the appropriate multi-hop transition for prediction. Therefore, we define \emph{substitutability} as a unique generalization type involving only multi-hop item transitions.

\begin{tcolorbox}[colback=GrayBlue,colframe=GrayBlue,arc=5pt,boxrule=0pt,left=2pt,right=2pt,top=0pt,bottom=0pt]
\textbf{Def. (Substitutability).} A data instance $(u, i_t) \notin \mathcal{D}_{\text{mem}}$ satisfies \textit{substitutability} if:
\begin{equation*}
    \exists k \ge 2, \exists u' \in \mathcal{D}_{\text{train}} \text{ s.t. } [i_{t-k} \to \dots \to i_t] \subseteq u'.
\end{equation*}
\end{tcolorbox}

\begin{table*}[t]
\centering
\caption{Performance breakdown by memorization or generalization categories. We report NDCG@10 for each dataset. \textbf{Bold} indicates the best performing model. ``Mem.'' stands for ``memorization'', and ``UC'' stands for ``uncategorized''. 
Note that a single data instance may fall into multiple generalization categories, so the sum of the ratios of the generalization subsets can exceed that of ``All''. In contrast, the proportions of memorization, generalization, and uncategorized instances sum to 100\%.}
\label{tab:generalization_full}

\small
\setlength{\tabcolsep}{3.5pt}
\renewcommand{\arraystretch}{1.08}

\adjustbox{max width=\textwidth}{
\begin{tabular}{ll c c ccc cccc cccc cccc c}
\toprule
\multicolumn{1}{l}{\multirow{4}{*}{\textbf{Dataset}}} & \multicolumn{1}{l}{\multirow{4}{*}{\textbf{Model}}} & \multirow{4}{*}{\textbf{Mem.}} & \multicolumn{16}{c}{\textbf{Generalization}} & \multirow{4}{*}{\textbf{UC}} \\
\cmidrule(lr){4-19} 
& & & \multirow{2}{*}{\textbf{All}} & \multicolumn{3}{c}{\textbf{Substitutability}} & \multicolumn{4}{c}{\textbf{Symmetry}} & \multicolumn{4}{c}{\textbf{Transitivity}} & \multicolumn{4}{c}{\textbf{2nd-Symmetry}} & \\
\cmidrule(lr){5-7} \cmidrule(lr){8-11} \cmidrule(lr){12-15} \cmidrule(lr){16-19} 
 &  & & & \textbf{2} & \textbf{3} & \textbf{4} & \textbf{1} & \textbf{2} & \textbf{3} & \textbf{4} & \textbf{1} & \textbf{2} & \textbf{3} & \textbf{4} & \textbf{1} & \textbf{2} & \textbf{3} & \textbf{4} & \\
\midrule

\rowcolor{GrayBlue}
\cellcolor{white}\multirow{3}{*}{\shortstack[c]{\textbf{Sports}\\ \textbf{(2014)}}}
& Ratio & 5.1 & 84.7 & 5.4 & 4.7 & 4.2 & 1.4 & 3.3 & 3.6 & 3.4 & 15.1 & 26.1 & 18.0 & 11.1 & 25.1 & 32.9 & 15.8 & 7.9 & 10.2 \\
& SASRec    & \textbf{.2816} & .0128 & .0732 & .0446 & .0279 & .0767 & .0556 & .0414 & .0368 & .0433 & .0120 & .0046 & .0016 & .0321 & .0072 & .0020 & .0012 & \textbf{.0004} \\
& TIGER & .1656 & \textbf{.0179} & \textbf{.1059} & \textbf{.0650} & \textbf{.0415} & \textbf{.0963} & \textbf{.0762} & \textbf{.0650} & \textbf{.0591} & \textbf{.0499} & \textbf{.0215} & \textbf{.0086} & \textbf{.0023} & \textbf{.0405} & \textbf{.0140} & .0015 & \textbf{.0013} & .0003 \\
\cmidrule{1-20}

\rowcolor{GrayBlue}
\cellcolor{white}\multirow{3}{*}{\shortstack[c]{\textbf{Beauty}\\ \textbf{(2014)}}}
& Ratio & 8.6 & 81.3 & 6.7 & 5.5 & 4.4 & 2.1 & 3.6 & 3.4 & 2.8 & 16.2 & 22.6 & 16.7 & 10.8 & 24.1 & 30.8 & 15.2 & 7.9 & 10.1 \\
& SASRec    & \textbf{.3793} & .0134 & .0814 & .0395 & .0235 & .0773 & .0535 & .0510 & .0391 & .0461 & .0105 & .0044 & .0026 & .0342 & .0076 & .0018 & .0008 & .0001 \\
& TIGER & .2456 & \textbf{.0210} & \textbf{.1046} & \textbf{.0648} & \textbf{.0381} & \textbf{.1156} & \textbf{.1026} & \textbf{.0696} & \textbf{.0437} & \textbf{.0551} & \textbf{.0235} & \textbf{.0113} & \textbf{.0059} & \textbf{.0468} & \textbf{.0167} & \textbf{.0029} & \textbf{.0015} & \textbf{.0015} \\
\cmidrule{1-20}

\rowcolor{GrayBlue}
\cellcolor{white}\multirow{3}{*}{\shortstack[c]{\textbf{Science}\\ \textbf{(2023)}}}
& Ratio & 4.0 & 84.2 & 4.6 & 4.1 & 3.5 & 2.0 & 3.7 & 3.5 & 2.9 & 16.5 & 24.8 & 17.2 & 10.8 & 26.3 & 30.7 & 15.4 & 8.3 & 11.8 \\
& SASRec    & \textbf{.2452} & .0132 & .0902 & .0510 & .0339 & \textbf{.1073} & .0722 & .0436 & .0346 & .0435 & .0118 & .0039 & .0021 & .0329 & .0070 & .0014 & .0009 & .0002 \\
& TIGER & .2348 & \textbf{.0177} & \textbf{.1114} & \textbf{.0688} & \textbf{.0447} & .1033 & \textbf{.1040} & \textbf{.0619} & \textbf{.0459} & \textbf{.0512} & \textbf{.0189} & \textbf{.0070} & \textbf{.0037} & \textbf{.0400} & \textbf{.0123} & \textbf{.0030} & \textbf{.0011} & \textbf{.0003} \\
\cmidrule{1-20}

\rowcolor{GrayBlue}
\cellcolor{white}\multirow{3}{*}{\shortstack[c]{\textbf{Music}\\ \textbf{(2023)}}}
& Ratio & 6.9 & 89.8 & 7.5 & 6.4 & 5.6 & 3.1 & 6.3 & 5.9 & 5.0 & 30.4 & 30.3 & 15.0 & 7.1 & 43.3 & 30.7 & 10.4 & 4.2 & 3.3 \\
& SASRec    & .1973 & .0173 & .0818 & .0513 & .0319 & \textbf{.0920} & .0678 & .0502 & .0294 & .0394 & .0095 & .0033 & .0013 & .0314 & .0057 & \textbf{.0011} & \textbf{.0011} & .0000 \\
& TIGER & \textbf{.2020} & \textbf{.0205} & \textbf{.0976} & \textbf{.0574} & \textbf{.0376} & .0879 & \textbf{.0814} & \textbf{.0581} & \textbf{.0396} & \textbf{.0434} & \textbf{.0144} & \textbf{.0043} & \textbf{.0014} & \textbf{.0359} & \textbf{.0087} & .0014 & .0007 & \textbf{.0004} \\
\cmidrule{1-20}

\rowcolor{GrayBlue}
\cellcolor{white}\multirow{3}{*}{\shortstack[c]{\textbf{Office}\\ \textbf{(2023)}}}
& Ratio & 4.4 & 86.3 & 4.5 & 4.1 & 3.6 & 1.9 & 3.7 & 3.4 & 2.9 & 22.4 & 27.4 & 16.4 & 9.5 & 33.1 & 30.1 & 13.5 & 6.8 & 9.3 \\
& SASRec    & .2405 & .0097 & .0682 & .0412 & .0282 & .0729 & .0612 & .0398 & .0273 & .0277 & .0062 & .0019 & .0012 & .0212 & .0037 & .0012 & .0006 & \textbf{.0002} \\
& TIGER & \textbf{.2719} & \textbf{.0154} & \textbf{.1092} & \textbf{.0667} & \textbf{.0433} & \textbf{.0914} & \textbf{.1025} & \textbf{.0649} & \textbf{.0470} & \textbf{.0384} & \textbf{.0136} & \textbf{.0040} & \textbf{.0020} & \textbf{.0311} & \textbf{.0089} & \textbf{.0017} & \textbf{.0007} & .0001 \\
\cmidrule{1-20}

\rowcolor{GrayBlue}
\cellcolor{white}\multirow{3}{*}{\textbf{Steam}}
& Ratio & 37.4 & 61.5 & 18.7 & 11.9 & 7.1 & 7.5 & 12.7 & 9.3 & 6.0 & 51.0 & 8.0 & 1.6 & 0.5 & 54.9 & 5.2 & 0.9 & 0.4 & 1.1 \\
& SASRec    & .3855 & .0133 & .0271 & .0129 & \textbf{.0092} & \textbf{.0306} & .0216 & .0139 & .0091 & .0144 & \textbf{.0079} & \textbf{.0079} & \textbf{.0112} & .0143 & .0057 & \textbf{.0018} & \textbf{.0051} & .0000 \\
& TIGER & \textbf{.3885} & \textbf{.0157} & \textbf{.0366} & \textbf{.0149} & .0073 & .0305 & \textbf{.0331} & \textbf{.0169} & \textbf{.0109} & \textbf{.0175} & \textbf{.0079} & .0038 & .0000 & \textbf{.0168} & \textbf{.0077} & .0007 & .0000 & .0000 \\
\cmidrule{1-20}

\rowcolor{GrayBlue}
\cellcolor{white}\multirow{3}{*}{\textbf{Yelp}}
& Ratio & 4.2 & 92.8 & 6.4 & 7.8 & 6.9 & 2.7 & 7.1 & 6.5 & 5.8 & 22.6 & 33.5 & 18.5 & 9.7 & 38.7 & 36.8 & 12.0 & 4.4 & 3.0 \\
& SASRec    & \textbf{.2487} & \textbf{.0233} & .0858 & \textbf{.0714} & .0435 & \textbf{.0724} & .0855 & \textbf{.0484} & .0300 & .0436 & \textbf{.0215} & \textbf{.0133} & \textbf{.0139} & \textbf{.0439} & .0120 & .0016 & .0010 & \textbf{.0007} \\
& TIGER & .1402 & .0213 & \textbf{.0898} & .0620 & \textbf{.0445} & .0659 & \textbf{.0861} & .0479 & \textbf{.0389} & \textbf{.0436} & .0213 & .0112 & .0055 & .0356 & \textbf{.0150} & \textbf{.0034} & \textbf{.0021} & .0000 \\
\bottomrule
\end{tabular}
}
\end{table*}

\subsection{Uncategorized Data Instance}\label{subsec:uncategorized}

Given a maximum hop count (\eg $4$ in our experiments, see~\Cref{sec:performance_breakdown}), any data instance that is neither memorization-related nor generalization-related is labeled as ``uncategorized.'' Such instances may involve items unseen during training, exhibit higher-order transition patterns, require capabilities beyond the scope of memorization and generalization, or be inherently unpredictable based on historical data alone. In our experiments, we also analyze model performance on these uncategorized instances.

\section{Performance Breakdown: Item IDs vs. Semantic IDs}
\label{sec:performance_breakdown}
In this section, we present the empirical results for memorization and generalization capabilities for GR and item ID-based models.

\subsection{Experiment Setup}
\paratitle{Datasets.}
We conduct experiments on seven public datasets that are widely used in evaluating GR models~\citep{rajput2023tiger,liu2025e2egrec,yang2025liger,wang2024letter}: ``Sports and Outdoors'' (\textbf{Sports}) and ``Beauty'' (\textbf{Beauty}) from the Amazon Reviews 2014 collection~\citep{mcauley2015amazon}; ``Industrial and Scientific'' (\textbf{Science}), ``Musical Instruments'' (\textbf{Music}), and ``Office Products'' (\textbf{Office}) from the Amazon 2023 collection~\citep{hou2024bridging}; \textbf{Steam}~\citep{kang2018sasrec} and \textbf{Yelp}\footnote{https://business.yelp.com/data/resources/open-dataset/}. 
The statistics of processed datasets have been reported in~\Cref{tab:dataset_statistics}.
We adopt the standard leave-last-out data split, using the last and second-to-last items of each sequence for testing and validation, respectively.

\begin{wraptable}{r}{0.40\linewidth}
    \centering
    \vspace{-15pt}
    \caption{Dataset statistics.}
    \label{tab:dataset_statistics}
    
    \scriptsize 
    \setlength{\tabcolsep}{3.5pt}
    \renewcommand{\arraystretch}{1.08}
    
    \adjustbox{max width=\linewidth}{
    \begin{tabular}{lrrrrr}
    \toprule
    \textbf{Dataset} & \textbf{Users} & \textbf{Items} & \textbf{Interactions} & \textbf{Sparsity} & \textbf{AvgLen} \\
    \midrule
    Sports  & 35,599  & 18,358 & 296,337   & 99.95\% & 8.32  \\
    Beauty  & 22,364  & 12,102 & 198,502   & 99.93\% & 8.88  \\
    Science & 50,986  & 25,849 & 412,947   & 99.97\% & 8.10  \\
    Music   & 57,440  & 24,588 & 511,836   & 99.96\% & 8.91  \\
    Office  & 223,309 & 77,552 & 1,800,878 & 99.99\% & 8.06  \\
    Steam   & 47,762  & 10,972 & 599,620   & 99.89\% & 12.55 \\
    Yelp    & 30,432  & 20,034 & 316,354   & 99.95\% & 10.40 \\
    \bottomrule
    \end{tabular}
    } 
    
    \vspace{-10pt}
\end{wraptable}

\paratitle{Models.}
We benchmark two models: TIGER~\citep{rajput2023tiger}, representing the generative recommendation paradigm, and SASRec~\citep{kang2018sasrec}, representing the conventional sequential recommendation paradigm. Note that, for a fair comparison, we optimize the SASRec model using cross-entropy loss and treat all items as negative samples, following~\citet{liu2025e2egrec}, rather than sampling a single negative item per instance as in~\citet{rajput2023tiger}.

\paratitle{Implementation details.}
We fine-tune the learning rate over $\{1\text{e-}3, 3\text{e-}3\}$ and train for a maximum of 150 epochs with early stopping. The checkpoint achieving the best validation performance is selected for testing.

\paratitle{Data categorization.}
We partition test instances into:
\textbf{memorization}, \textbf{generalization}
(if memorization is not satisfied), and \textbf{uncategorized} (if none of the above is satisfied).
Note that the above three categories are mutually exclusive.
However, a data instance may exhibit multiple generalization types, each associated with several possible hop distances. Following Occam's razor, we annotate an instance with all applicable types but retain only the minimum hop distance for each type.


\subsection{Performance Analysis}

In this section, we analyze the performance comparison between SASRec and TIGER, broken down by the memorization and generalization categories defined in~\Cref{sec:defining-memorization-and-generalization}.

\paratitle{SASRec memorizes, TIGER generalizes.}
As illustrated in~\Cref{tab:generalization_full}, TIGER generally underperforms SASRec on memorization subsets
(\eg \textcolor{PerfRed}{$\boldsymbol{-43.6\%}$} on \textbf{Yelp},
\textcolor{PerfRed}{$\boldsymbol{-41.2\%}$} on \textbf{Sports},
\textcolor{PerfRed}{$\boldsymbol{-35.2\%}$} on \textbf{Beauty}, and comparable on others),
while consistently outperforming SASRec on generalization subsets
(\eg \textcolor{PerfGreen}{$\boldsymbol{+58.8\%}$} on \textbf{Office},
\textcolor{PerfGreen}{$\boldsymbol{+56.7\%}$} on \textbf{Beauty},
and \textcolor{PerfGreen}{$\boldsymbol{+39.8\%}$} on \textbf{Sports}).
This trade-off suggests that SASRec relies more on memorizing observed patterns, while TIGER is more effective at composing learned item transitions for generalization.
Both models achieve substantially higher performance on memorization than on generalization overall, reflecting the intrinsic difficulty of generalizing beyond observed transitions.
Moreover, both models exhibit near-zero performance on the uncategorized subset while achieving reasonable performance on the others. This supports the validity of our data attribution and suggests that the uncategorized instances are indeed difficult to predict, consistent with our hypothesis in~\Cref{subsec:uncategorized}.

\paratitle{Generalization categories.}
Comparing performance across generalization categories, we observe that both models achieve higher performance on \textit{Substitutability} and \textit{Symmetry} than on \textit{Transitivity} and \textit{2nd-Symmetry}.
We attribute this to differences in the difficulty of the various generalization types. Substitutability and Symmetry require induction from only a single training example, whereas Transitivity and second-order Symmetry require composing knowledge from multiple examples, representing a structurally more complex form of generalization.

\paratitle{Generalization hops.}
Within each generalization category, both models perform monotonically worse as hop distance increases. This shows that nearby item transitions pose a stronger influence than distant ones.
In low-hop settings, SASRec can sometimes outperform TIGER. But its performance drops faster as the hop distance grows.
The decline is even sharper in more difficult categories such as Transitivity and 2nd-Symmetry.
This suggests that SASRec mainly generalizes over local context, while TIGER remains more robust for longer-hop generalization.

\paratitle{Data ratio analysis.}
Finally, we examine the proportion of test instances in each category.
In all datasets, memorization cases form a much smaller portion than generalization cases. This suggests that pure memorization is limited, and effective recommendation requires substantial generalization capability. 
Among the generalization categories, most instances require combining information from multiple training examples, whereas only a small fraction can be inferred from a single training instance (Substitutability and Symmetry).
Uncategorized instances consistently account for less than $10\%$ of the data, indicating that most test transitions can be explained by other categories.


\section{Mechanism Analysis: A Token-Level Lens}
\label{sec:token_mem}

Semantic ID-based GR models generally outperform item ID-based models on generalization-related subsets (\Cref{sec:performance_breakdown}). This raises a question: \emph{Why does GR generalize better yet memorize worse than item ID-based models?} In this section, we investigate the underlying mechanisms of GR models through a token-level lens. 

We first introduce the concept of \emph{prefix n-gram memorization} (Section~\ref{sec:token-mem-def}), and demonstrate that item-level generalization can often be interpreted as token-level memorization within the semantic ID space (Section~\ref{sec:reduce_item_to_token}). Next, we characterize models' behavior using this new lens (Section~\ref{sec:performance-tradeoff-analysis}), and find that:
(1) GR generalization performance improves when the underlying token transitions are more frequently observed in the training data. (2) Different item transitions can share the same memorized prefix, which can decrease GR's ability to memorize a specific item transition. Finally, to further validate our hypothesis, we design a controlled study to vary the token memorization ratio and measure its direct impact on the generalization-memorization trade-off (Section~\ref{subsec:mechanism-validation}).


\subsection{Prefix N-Gram Memorization}
\label{sec:token-mem-def}

\begin{wrapfigure}{r}{0.50\linewidth}
    \vspace{-10pt}
    \centering
    \includegraphics[width=\linewidth]{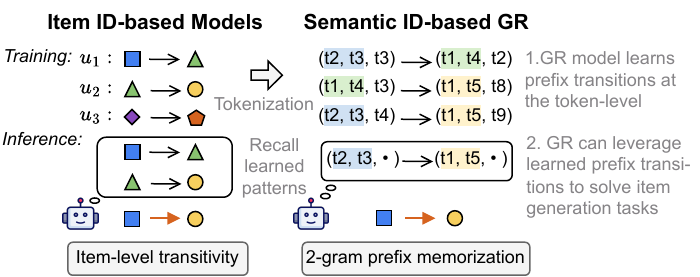}
    \caption{Illustration of how item-level generalization can be reduced to token-level memorization for GR models.}
    \label{fig:prefix-mem-mechanism}
    \vspace{-10pt}
\end{wrapfigure}

\paratitle{Motivation.} Unlike item ID-based models, GR models represent items as sequences of discrete semantic ID tokens shared across items. This allows the model to anchor predictions on sub-item level transition patterns.
However, quantifying memorization behavior at a token level is non-trivial. Directly attributing the effect of a single token on another token is difficult because token-to-token correlations are dense and highly dependent on the contexts~\citep{grosse2023studying}. For LLMs, people often assess memorization via $n$-gram correlation between context and target text, reflecting the model's ability to memorize the n-gram `knowledge' and generate the corresponding n-gram `answer'~\citep{liu2401infini,wang2024generalization}. Drawing inspiration from this, we propose quantifying memorization in GR by considering the \emph{prefix $n$-grams} of the context-target item pair. Since semantic IDs encode hierarchical (coarse-to-fine) semantic information, focusing on transitions from one prefix to another prefix captures the most prominent semantic dependencies (\Cref{fig:prefix-mem-mechanism}).

\paratitle{Token Prefix.}
Let $\mathrm{tok}(i) = [z_1, z_2, \ldots, z_L]$ denote the semantic-ID tokenization of item $i$, where $L$ is the number of tokens in the semantic ID.
For a prefix length $n \le L$, define the $n$-gram prefix operator
$
\mathrm{pref}_n(i) \triangleq [z_1,\ldots,z_n].
$


\paratitle{Prefix n-gram memorization.} 
We define token-level memorization by considering only the semantic ID prefixes of items in the transitions. A test instance is considered \emph{$n$-gram prefix-memorizable} if the first $n$ tokens (the $n$-gram prefix) of both items in the target transition $[i_s\rightarrow i_t]$ occur in the training set, even when the exact items differ. 

\begin{tcolorbox}[colback=GrayBlue,colframe=GrayBlue,arc=5pt,boxrule=0pt,left=2pt,right=2pt,top=0pt,bottom=0pt]
\textbf{Def. (1-hop Prefix N-Gram Memorization).}
A data instance $(u, i_t)$ satisfies \textit{$1$-hop prefix $n$-gram memorization} if:
\begin{equation*}
\begin{aligned}
\exists\, u' \in \mathcal{D}_{\text{train}},\ \exists\, s \ge 2 \ \text{s.t.}\quad
& [j_{s-1} \to j_s] \subseteq u', \\
& \mathrm{pref}_n(i_{t-1}) = \mathrm{pref}_n(j_{s-1}), \\
& \mathrm{pref}_n(i_t) = \mathrm{pref}_n(j_s).
\end{aligned}
\end{equation*}
\end{tcolorbox}

Analogous to the multi-hop generation framework, this definition naturally extends to $k$-hop transitions. Notably, when $k = 1$, token prefix memorization can be viewed as a relaxed form of \emph{memorization}, whereas for $k > 1$, it is analogous to the definition of \emph{substitutability} (see~\Cref{subsec:multi-hop-gen}). For the following sections, we refer to prefix $n$-gram memorization as \emph{token memorization} for brevity.

\subsection{From Item Generalization to Token Memorization}
\label{sec:reduce_item_to_token}

\begin{figure}[b]
    \centering
    \includegraphics[width=1.0\columnwidth]{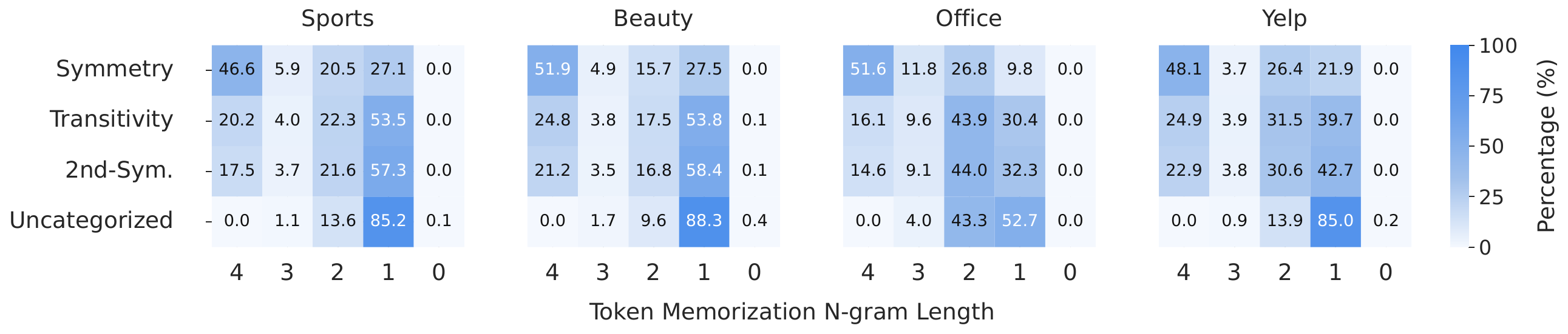}
    \caption{Token memorization ratios for each item-level generalization category. The X-axis represents the prefix length for token memorization.}
    \label{fig:token-mem-conversion}
\end{figure}

We examine how item-level generalization can be reduced to token memorization for GR models, following the definition in \Cref{sec:token-mem-def}. Unless otherwise specified, the following experiments on all datasets aggregate all token memorizations with $hop \leq 4$, and utilize a $256 \times 3$ semantic ID quantization followed by one identifier token, consistent with~\citet{rajput2023tiger}.

\paratitle{Item generalization instances often reduce to token memorization for GR models.} Figure~\ref{fig:token-mem-conversion} illustrates the reduction of item-level generalization categories into token-level prefix-gram memorization. 
We observe that a non-trivial fraction of instances reduce to 1-, 2-, and 3-gram prefix memorization. 
For example, on average, more than 5\% of item-level generalization transitions (symmetry, transitivity, and 2nd-symmetry) can also be explained as 3-hop prefix memorization. 
Notably, the vast majority of test instances ($>99\%$) across all item-level categories admit at least 1-gram prefix memorization. This demonstrates that for many test instances where the item-level transition is unseen, the training set nevertheless contains matching prefix transitions, allowing the model to leverage prefix memorization for inference.



\paratitle{Token memorization ratio reflects item-level difficulty.} Across categories, symmetry exhibits a higher ratio of 4-gram memorization, largely due to its overlap with item-level substitutability. In contrast, transitivity and 2nd-order symmetry mostly reduces to short prefix memorization (2–3 grams), yielding weaker prefix-transition support from training and making these tasks harder. Furthermore, uncategorized instances reduce almost exclusively to 1-gram memorization, representing the weakest form of prefix-transition support. Overall, these findings show that the ratio of token memorization directly reflects the item-level task difficulty and the model performance trends observed in~\Cref{sec:performance_breakdown}.




\subsection{Explaining Performance Trade-off via Token Memorization}
\label{sec:performance-tradeoff-analysis}

Having established that item-level generalization often reduces to token-level memorization, we now investigate whether this mechanism explains the performance trade-off: GR generalizes better but memorizes worse at item level. We categorize test instances through token memorization and show that \emph{token memorization enables better generalization} (\Cref{subsubsec:better-generalization}), and \emph{token memorization dilutes item memorization} (\Cref{subsubsec:dilute-memorization}).


\begin{figure}[t]
    \centering
    \begin{minipage}[t]{0.44\linewidth}
        \centering
        \includegraphics[width=\linewidth]{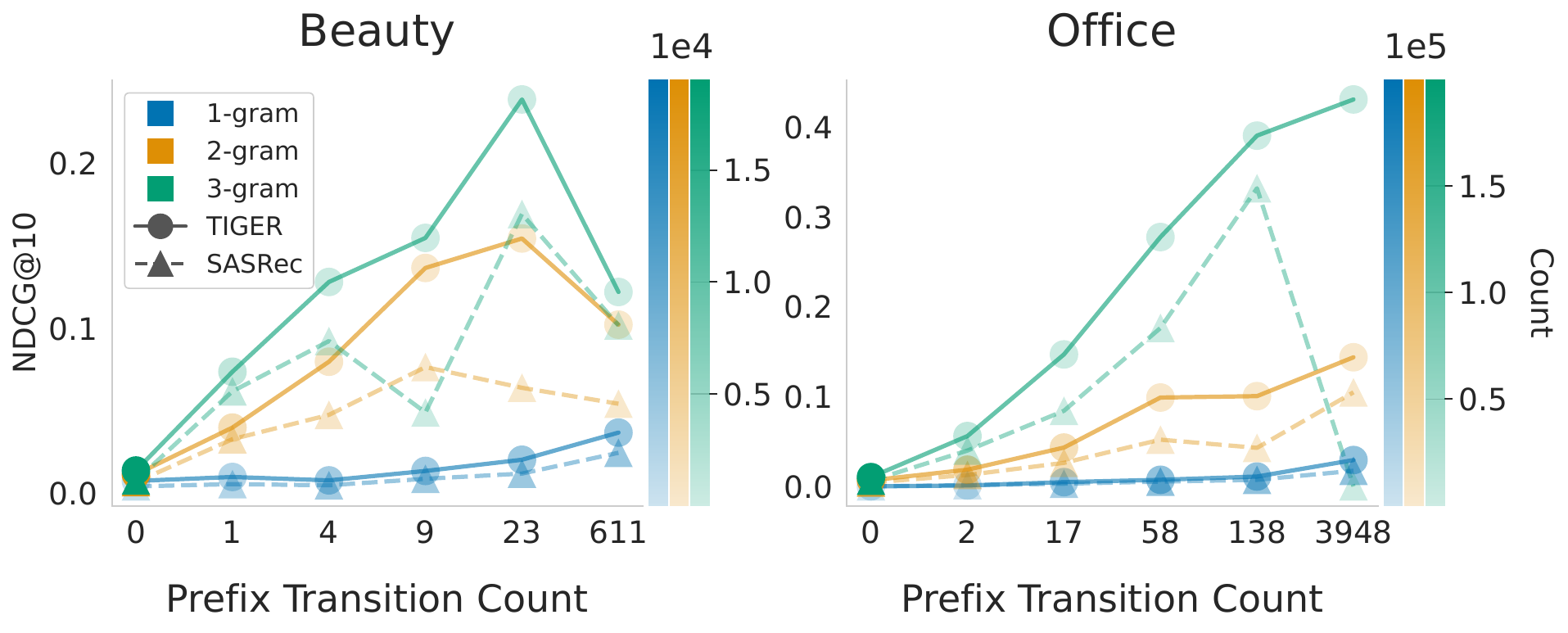}
        \captionof{figure}{Test NDCG@10 against prefix transition counts on 2 datasets. Transition counts are grouped by quantiles.}
        \label{fig:performance-vs-prefix-combined}
    \end{minipage}\hfill
    \begin{minipage}[t]{0.54\linewidth}
        \centering
        \includegraphics[width=\linewidth]{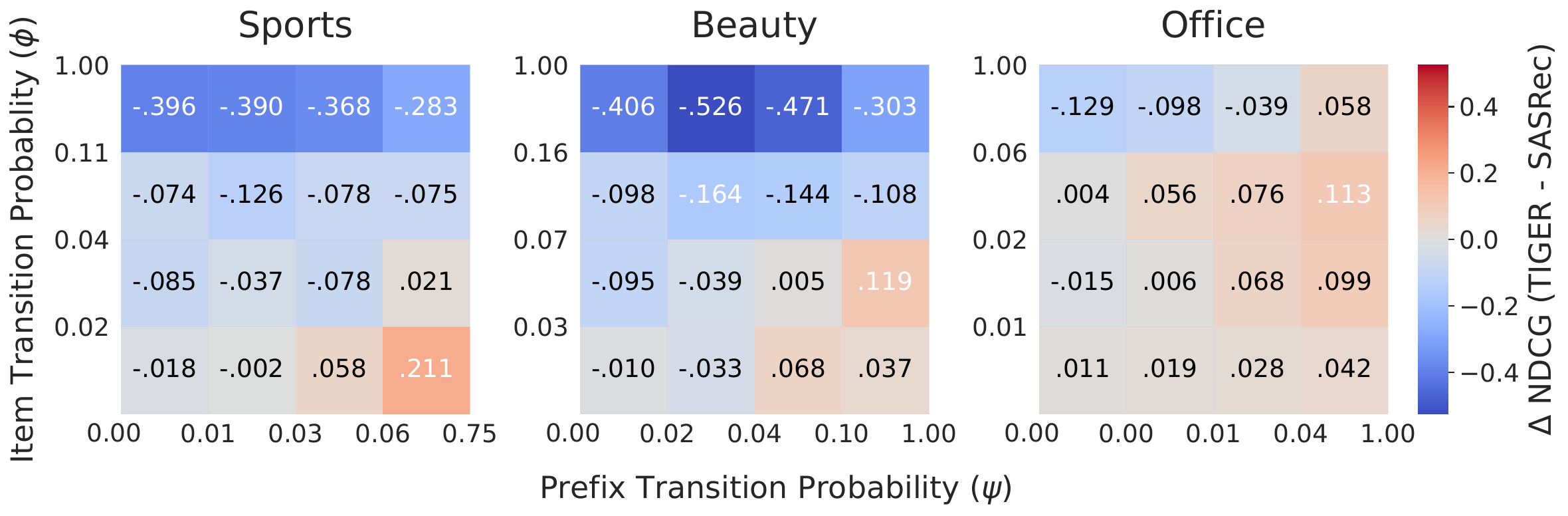}
        \captionof{figure}{TIGER Advantage breakdown by prefix transition probability and item transition probability. Both axes are grouped by quantiles.}
        \label{fig:heatmap-comparison}
    \end{minipage}
\end{figure}

\subsubsection{Token Memorization Enables Better Generalization}
\label{subsubsec:better-generalization}
To analyze how TIGER's performance correlates with token memorization support at a finer granularity, we quantify prefix support as the count of prefix-transition occurrences in the training data:
\[
C_n(i_{t-k}, i_t) = C(\mathrm{pref}_n(i_{t-k}) \to \mathrm{pref}_n(i_t)),
\]
where $C(\cdot)$ is the count function. Given an item-level generalization instance $(u, i_t)$, the token memorization support $C_n(u, i_t) = \sum_{k=1}^{K} C_n(i_{t-k}, i_t)$ is the sum of prefix supports of all the k-hop transitions to the target item.

\paratitle{More token memorization support correlates with better generalization}. We aggregate token memorization support for hop $k \leq4$, and plot the TIGER and SASRec models' NDCG@10 on all item generalization instances (without item memorization support) against $C_n(u, i_t)$
in~\Cref{fig:performance-vs-prefix-combined}. First,
we observe that supported instances consistently achieve substantially higher NDCG@10 than non-supported instances for both TIGER and SASRec. TIGER’s gain over SASRec is substantially larger on prefix-supported instances, indicating that TIGER benefits from prefix-transition support that SASRec does not explicitly model.Furthermore, TIGER's performance gap increases with both increasing support count $C_n(u, i_t)$ and prefix length $n$, indicating that more token memorization support correlates with better generalization performance.

\subsubsection{Token Memorization Dilutes Item Memorization}
\label{subsubsec:dilute-memorization}
Conversely, we investigate why this same mechanism might degrade performance on item memorization tasks. We hypothesize a dilution effect: while SASRec directly optimizes the specific item transition $i_{t-1} \to i_{t}$, TIGER predicts through prefix transitions that are shared by many items. Thus, its ability to memorize specific item transitions can be diluted. To verify this, we evaluate TIGER's relative performance ($\Delta \text{NDCG}$) against two metrics: \emph{item transition probability} ($\phi$), the probability of predicting target item based on all item transitions, and \emph{prefix transition probability} ($\psi$), the probability of predicting target item's prefix based on all prefix transitions. For a transition $i_{t-1} \to i_t$, these quantities are estimated empirically by:
\[
\phi = \frac{C(i_{t-1} \to i_t)}{C(i_{t-1} \to \cdot)}, \quad \psi = \frac{C(\text{pref}_n(i_{t-1}) \to \text{pref}_n(i_t))}{C(\text{pref}_n(i_{t-1}) \to \cdot)},
\]
where $C$ is the count function and $C(x \rightarrow \cdot)$ is the total number of outgoing transitions from the context $x$.

\paratitle{Token memorization dilutes item memorization.} We visualize how TIGER’s NDCG gain over SASRec (\ie $\Delta \text{NDCG}$) breaks down by item transition probability $\phi$ and prefix transition probability $\psi$. As shown in~\Cref{fig:heatmap-comparison}, across all datasets, we observe significant NDCG loss when instances have high $\phi$ but low $\psi$. 
In this case, TIGER spreads probability mass across many items sharing the same prefix transition, causing difficulty for the model to memorize a specific item transition pattern.
Conversely, TIGER can even outperform SASRec when $\psi$ is high, indicating that TIGER’s item memorization performance is strong only when token memorization aligns. Across the datasets, token memorization usually leads to more dilution cases, causing TIGER to
underperform SASRec on memorization tasks.

\begin{figure}[t]
    \centering
    \begin{minipage}[t]{0.45\columnwidth}
        \vspace{0pt}
        \centering
        \captionof{table}{Experiment configurations and token memorization ratio across different semantic ID (SID) configurations.}
        \setlength{\tabcolsep}{3.8pt}
        \renewcommand{\arraystretch}{1.05}
        \resizebox{\linewidth}{!}{%
        \begin{tabular}{lcrcccccc}
        \toprule
        \multicolumn{2}{c}{\textbf{SID Configuration}} & \textbf{Budget} & \multicolumn{6}{c}{\textbf{Token Memorization Ratio (\%)}} \\
        \cmidrule(lr){1-2} \cmidrule(lr){3-3} \cmidrule(lr){4-9}
        Len ($L$) & Size ($V$) & TFLOPs & $n=5$ & $n=4$ & $n=3$ & $n=2$ & $n=1$ & $n=0$ \\
        \midrule
        \multirow{2}{*}{$L=2$} 
         & 1024 & $5.00 \times 10^{15}$ & -- & -- & -- & 15.06 & 84.47 & 0.47 \\
         & 4096 & $5.00 \times 10^{15}$ & -- & -- & -- & 15.06 & 83.93 & 1.01 \\
        \midrule
        \multirow{2}{*}{$L=3$} 
         & 256  & $7.50 \times 10^{15}$ & -- & -- & 15.06 & 16.57 & 68.36 & 0.01 \\
         & 4096 & $7.50 \times 10^{15}$ & -- & -- & 15.06 & 5.40  & 79.54 & 0.01 \\
        \midrule
        \multirow{2}{*}{$L=4$} 
         & 64   & $1.00 \times 10^{16}$ & -- & 15.06 & 7.68 & 46.26 & 31.01 & 0.00 \\
         & 1024 & $1.00 \times 10^{16}$ & -- & 15.06 & 0.92 & 21.34 & 62.68 & 0.00 \\
        \midrule
        \multirow{2}{*}{$L=5$} 
         & 64   & $1.33 \times 10^{16}$ & 15.06 & 1.22 & 5.74 & 46.08 & 31.89 & 0.00 \\
         & 256  & $1.33 \times 10^{16}$ & 15.06 & 0.20 & 1.09 & 16.69 & 66.96 & 0.00 \\
        \bottomrule
        \end{tabular}%
        }
        \label{tab:sid_conversion}
    \end{minipage}\hfill
    \begin{minipage}[t]{0.50\columnwidth}
        \vspace{0pt}
        \centering
        \includegraphics[width=\linewidth]{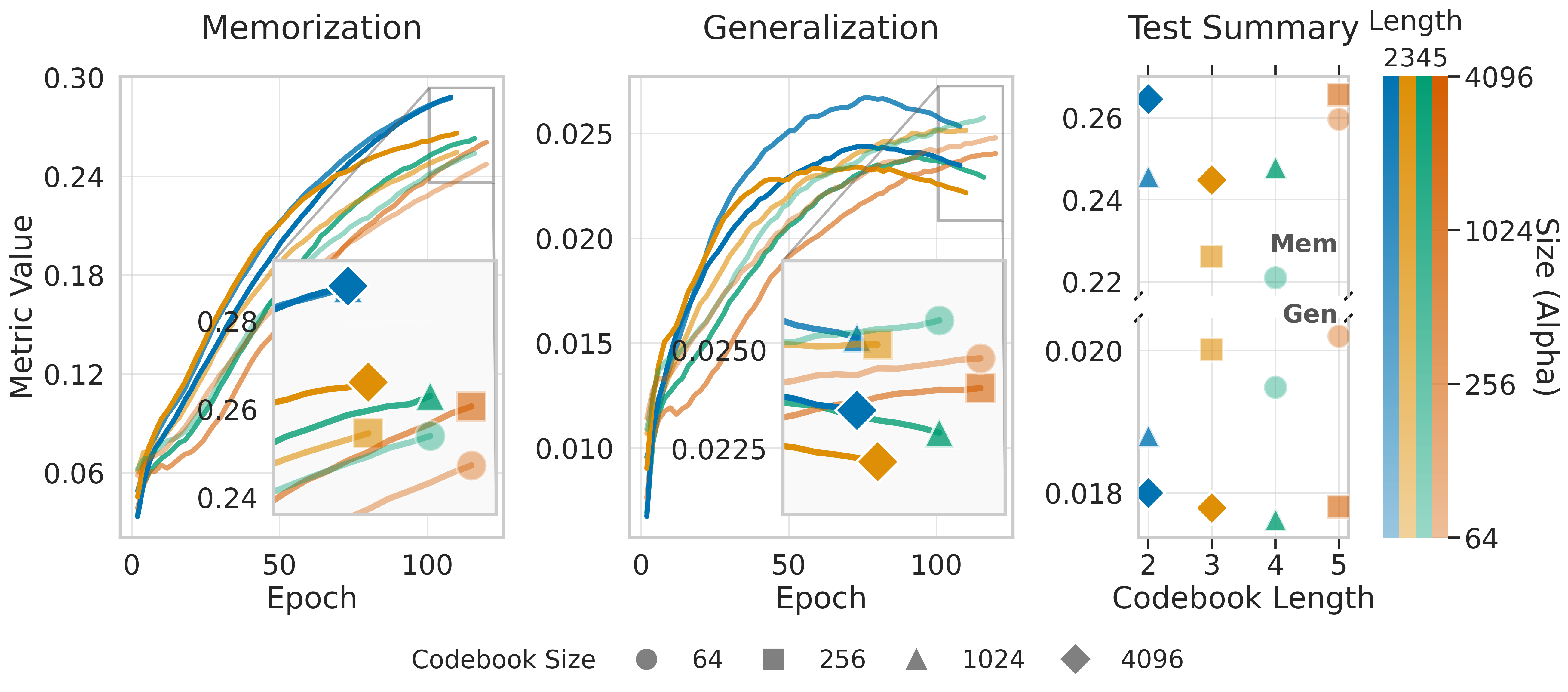}
        \captionof{figure}{Left/Middle: Val NDCG@10 training dynamics on the Memorization and Generalization subsets. Right: Test NDCG@10 for different codebook configurations.}
        \label{fig:mechanism-validation}
    \end{minipage}
\end{figure}

\subsection{Mechanism Validation}
\label{subsec:mechanism-validation}

In this section, we test our hypothesis that increasing token memorization ratio improves item-level generalization but can reduce item-level memorization. Our goal is to vary the token memorization ratio and measure the resulting changes in performance.

\paratitle{Experiment Setup.}
To manipulate the token memorization ratio in a systematic way, we vary the \emph{codebook size}. The intuition is that denser codebooks (smaller codebook sizes $V$) induce more prefix sharing across items, which increases the total number of token memorization instances. We consider SID lengths $L \in \{2,3,4,5\}$ (including identifier tokens), and for each fixed $L$ we evaluate two codebook sizes $V$. In~\Cref{tab:sid_conversion} we report the resulting token memorization ratios at each prefix length; as expected, smaller $V$ yields a larger fraction of instances with prefix-transition support.

\paratitle{Controlled factors.}
We only compare configurations \emph{within the same SID length} $L$, since optimizing different lengths can have different difficulty and longer SID length often leads to optimization issues~\citep{hou2025rpg}. We also fix the recommendation model size and match the training compute budget within each $L$. For each $L$, we estimate the training compute to ensure convergence, defined as both memorization and generalization metrics no longer increasing on the validation set. We report (i) validation training dynamics of NDCG@10 and (ii) final test NDCG@10 for each configuration.

\paratitle{Smaller codebook sizes lead to better generalization but worse memorization.}
From~\Cref{fig:mechanism-validation}, we observe a consistent pattern: smaller codebook size $V$ (denser codebooks, higher token memorization ratio) leads to better item-level generalization, but degraded memorization performance. Averaged across settings, the smaller $V$ improves generalization by 
\textcolor{PerfGreen}{$\boldsymbol{+10.24\%}$}
relative compared to the larger variant, while memorization decreases by 
\textcolor{PerfRed}{$\boldsymbol{-7.62\%}$}
relative. This provides consistent evidence across all tested codebook sizes that increasing token memorization ratio improves generalization while degrading memorization.

\paratitle{Regularization effect in training dynamics.}
From~\Cref{fig:mechanism-validation}, we further observe that for large $V$, generalization performance peaks early and then degrades as training proceeds. In contrast, for small $V$, generalization remains stable or continues to improve until convergence. This suggests that denser tokenization has a stronger data-level regularization effect. It encourages TIGER to rely more on shared prefix-transition structure, rather than overfitting to noisier, item-specific transitions.

\section{Memorization-Aware Adaptive Ensemble}


So far, we have shown that GR models excel on instances requiring generalization, whereas item ID-based models perform better on memorization.
Motivated by this observation, we investigate whether it is possible to combine the strengths of both paradigms.
We propose a hybrid framework that dynamically weights the predictions of a GR model and an item-based model for each given input, depending on the primary capability the input may require. We first introduce a simple, training-free indicator to estimate the likelihood of memorization (\Cref{subsec:routing_indicators}), followed by an empirical validation of their correlation with ground-truth categories (\Cref{subsec:indicator-validation}). Finally, we present experimental results demonstrating that this adaptive model ensemble strategy 
achieves improved overall performance (\Cref{subsec: routing-performance}).

\subsection{Adaptive Ensemble Indicators}
\label{subsec:routing_indicators}

The core idea of our adaptive ensemble strategy is to estimate how likely a data instance can be solved via memorization, and use this estimate to adjust the relative weights between the item ID-based model and the GR model for each instance. A key challenge is that the memorization criterion defined in~\Cref{subsec:mem} cannot be directly applied at inference time, since the target item is unavailable. We therefore introduce a practical indicator that estimates the memorization likelihood of a data instance $(u, i_t)$ and converts it into an ensemble weight.

\paratitle{Confidence-based indicator.} 
The intuition behind this approach is that memorization-related instances tend to lie closer to the training data distribution. As a result, an item ID-based model that effectively memorizes training patterns should produce more confident predictions on such instances. We therefore use the prediction confidence of the item ID-based model as a proxy for memorization likelihood. While confidence is often defined as the distance between prediction logits and a uniform distribution, the recommendation task typically involves a very large item space where most items receive negligible probability. We thus adopt the maximum softmax probability (MSP)~\citep{vashurin2025benchmarking} as a practical confidence score, defined as:
\begin{equation}
    s_{\text{Conf}}(u) = \max_{j \in \mathcal{I}} P_{\text{ID}}(i_t = j \mid u),
\end{equation}
where $P_{\text{ID}}(\cdot \mid u)$ denotes the predicted probability distribution of the ID-based model given user history $u$. We then transform $s_{\text{Conf}}(u)$ into an ensemble weight that controls the contribution of each model:
\begin{equation}
    \alpha(u) = \operatorname{sigmoid}(-q(s_{\text{Conf}}(u) - \tau)),\label{eqn:alpha}
\end{equation}
where $q$ and $\tau$ are hyperparameters.

\subsection{Validation of the Indicator}
\label{subsec:indicator-validation}

In this section, we empirically validate whether the proposed MSP indicator provides a meaningful estimate of memorization likelihood and can therefore support our adaptive ensemble strategy. Using the same datasets and models as in~\Cref{sec:performance_breakdown}, we group data instances into bins according to quantiles of their indicator values. For each bin, we examine both the proportion of memorization-related instances and the performance of the two paradigms.

\paratitle{Correlation with memorization.} 
As the indicator value increases, the proportion of memorization-related instances rises monotonically. This trend suggests that the indicator provides a meaningful estimate of memorization likelihood.

\paratitle{Performance crossover.} 
We also observe a performance crossover between the generative model TIGER and the ID-based model SASRec as the indicator value increases. At lower indicator values, TIGER achieves better performance, whereas at higher values SASRec becomes increasingly competitive and may outperform TIGER. This crossover pattern is consistent with our hypothesis: instances with high indicator values tend to be memorization-related and are better handled by ID-based models, while instances with low values favor the stronger generalization capability of GR models.

\begin{figure}[t]
    \centering
    \includegraphics[width=0.9\columnwidth]{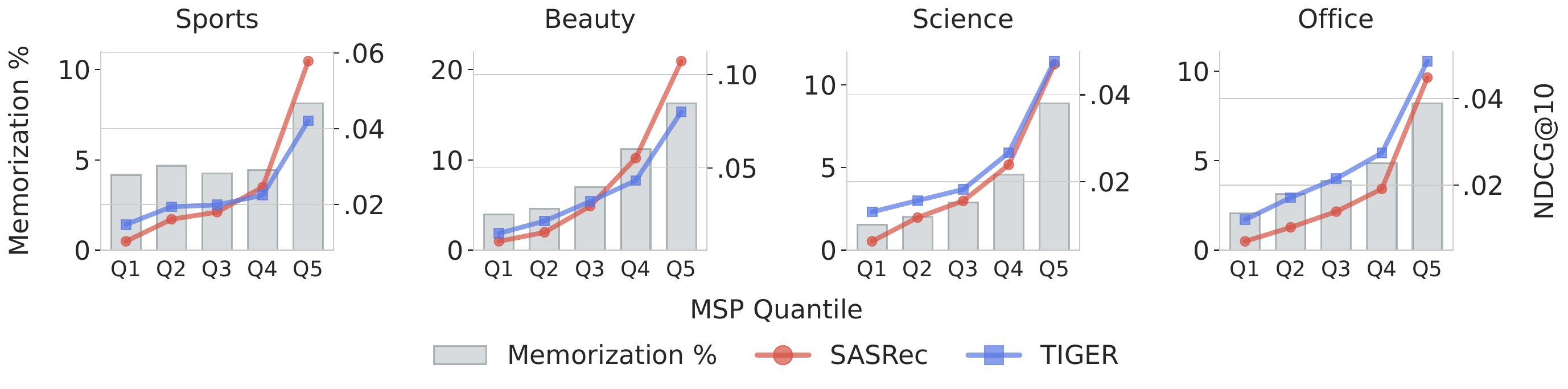}
    \caption{Correlation between the MSP indicator, proportion of memorization-related instances (bars), and model performance of SASRec and TIGER (lines).}
    \label{fig:indicator_correlation}
\end{figure}

\begin{table*}[t]
\centering
\footnotesize
\setlength{\tabcolsep}{4.5pt}
\renewcommand{\arraystretch}{1.08}
\caption{Performance comparison of ensemble strategies. The best performance is bolded.}
\label{tab:routing_performance}
\scalebox{0.95}{
\begin{tabular}{llccccccc}
\toprule
\textbf{Metric} & \textbf{Method} & \textbf{Sports} & \textbf{Beauty} & \textbf{Scientific} & \textbf{Instruments} & \textbf{Office} & \textbf{Steam} & \textbf{Yelp} \\
\midrule
\multirow{4}{*}{\textbf{N@10}}
& SASRec       & 0.0253 & 0.0436 & 0.0209 & 0.0291 & 0.0190 & 0.1525 & 0.0321 \\
& TIGER        & 0.0237 & 0.0383 & 0.0243 & 0.0323 & 0.0254 & 0.1551 & 0.0257 \\
& Fixed-weight & 0.0291 & 0.0471 & 0.0260 & 0.0343 & 0.0261 & \textbf{0.1579} & 0.0351 \\
& Adaptive     & \textbf{0.0296} & \textbf{0.0476} & \textbf{0.0261} & \textbf{0.0344} & \textbf{0.0264} & \textbf{0.1579} & \textbf{0.0352} \\
\midrule
\multirow{4}{*}{\textbf{R@10}}
& SASRec       & 0.0318 & 0.0566 & 0.0262 & 0.0319 & 0.0241 & 0.1738 & 0.0362 \\
& TIGER        & 0.0318 & 0.0542 & 0.0303 & 0.0357 & 0.0306 & 0.1773 & 0.0281 \\
& Fixed-weight & 0.0530 & 0.0838 & 0.0483 & 0.0635 & 0.0456 & \textbf{0.2008} & 0.0630 \\
& Adaptive     & \textbf{0.0537} & \textbf{0.0841} & \textbf{0.0485} & \textbf{0.0638} & \textbf{0.0461} & \textbf{0.2008} & \textbf{0.0637} \\
\bottomrule
\end{tabular}
}
\end{table*}

\subsection{Performance of Adaptive Ensemble}
\label{subsec: routing-performance}

Having validated the effectiveness of the proposed MSP indicators, we proceed to evaluate the overall performance of the memorization-aware adaptive model ensemble strategy on real-world datasets.


\paratitle{Experimental settings.} We use the same datasets and base models as in~\Cref{sec:performance_breakdown}. 
We use~\Cref{eqn:alpha} to map the raw MSP values to an instance-specific ensemble weight $\alpha(u) \in [0, 1]$. The final prediction is obtained by combining the scores produced by the two base models, SASRec and TIGER, according to this weight.
We compare our adaptive ensemble against the individual base models and a \textit{Fixed-weight Ensemble} baseline, which assigns a global constant weight $\alpha_{\text{static}}$ to SASRec for all instances. Hyperparameters are tuned on the validation set, with $q \in \{1,5,9,13\}$, $\tau \in \{0, 0.1, \dots, 0.5\}$, and $\alpha_{\text{static}} \in \{0, 0.1, \dots, 1.0\}$.

\paratitle{Recommendation results.} We present the performance comparison in~\Cref{tab:routing_performance} and make the following observations:
\begin{itemize}[leftmargin=*]
    \item The proposed adaptive ensemble strategy generally outperforms both the individual base models and the fixed-weight ensemble baseline. These results provide empirical evidence that item ID-based models and GR models offer complementary strengths.
    \item The improvement over baselines is more prominent on datasets with a stronger model cross-over effect (see~\Cref{fig:indicator_correlation}). The greater the comparative advantage each base model holds in its respective domain, the more the memorization-aware indicator can leverage this specialization.
\end{itemize}

\section{Related Works}
\label{sec:2-related}

\paratitle{Generative recommendation.} 
Unlike conventional sequential recommendation paradigms~\citep{rendle2010fpmc,kang2018sasrec,sun2019bert4rec,hidasi2016gru4rec,tang2018caser,ma2019hgn}, where each item is typically indexed by a unique ID, generative recommendation tokenizes each item as a sequence of sub-item tokens~\citep{tay2022dsi,rajput2023tiger,zhai2024hstu,deng2025onerec,he2025plum,zheng2024lcrec,www25-gen-rec-tutorial}. A generative model then operates over these token sequences and autoregressively generates tokens that index the next item. Existing work on generative recommendation primarily focuses on developing improved tokenization methods, such as exploring alternative algorithms~\citep{wang2024letter,zhu2024cost,hou2025rpg,hua2023p5cid,jin2024lmindexer,liu2025e2egrec,zhang2025psid,zheng2025utgrec,zheng2025mtgrec} or incorporating additional data sources~\citep{liu2024mbgen,hou2025actionpiece,zhu2025beyond,liu2024mmgrec,wang2025generative,zhong2025pctx}. Despite their empirical success, there remains a lack of systematic research examining which data instances generative recommendation outperforms conventional models on. Existing work typically attributes the strong performance of GR to semantic-integrated indexing systems~\citep{zhang2024moc,hou2025actionpiece,ju2025generative,liu2025gr_scaling} or finer-grained learning objectives~\citep{wang2024letter,zhu2024cost,hou2025rpg}. These studies offer design-centric explanations but lack a systematic analysis of how such designs translate into distinct prediction behaviors compared to conventional models. In this work, we conduct an in-depth comparative study of these two paradigms through the lens of memorization versus generalization, specifically aiming to answer the question: do GR models outperform conventional baselines because they generalize better?

\paratitle{Memorization and generalization.} 
The study of model memorization and generalization has long been a central concern in machine learning, spanning from classical machine learning models~\citep{han2022images,yang2023resmem,buchanan2025edge} to modern large language models~\citep{carlini2023quantifying,wang2024generalization,jiang2024investigating}. A key challenge of conducting these analyses is how to characterize the influence of training data on model behavior. One line of work approaches this question through \emph{counterfactual memorization}, which measures the causal effect of removing individual training instances on task performance~\citep{zhang2023counterfactual,grosse2023studying,raunak2021curious,ghosh2025rethinking}. However, these methods are often computationally expensive and difficult to scale~\citep{grosse2023studying,pruthi2020estimating,xia2024less}.
Other approaches define generalization as robustness to local input perturbations (\eg input clues~\citep{djire2025memorization} or task rules~\citep{xie2410memorization,barron2025too,chu2025sft}). In this work, we aim to investigate how generative recommendation models differ from conventional item ID-based models in their memorization and generalization capabilities. Because the recommendation task provides a concrete and well-defined ground-truth target, we categorize each data instance according to the primary capability required to make a correct prediction.

\section{Conclusion}

In this work, we presented a systematic study comparing semantic ID-based generative recommendation models with traditional item ID-based models through the lens of memorization and generalization. We proposed a framework to categorize data instances based on the item transition patterns they contain. Extensive experiments across various datasets revealed that generative recommendation models excel on generalization-related instances, while item ID-based models perform better on memorization-related instances. To explain this, we analyzed the models at the token level, finding that item-level generalization in generative models often reduces to token-level memorization. Finally, we show that the two paradigms are complementary by demonstrating that adaptively adjusting ensemble weights using a memorization-aware indicator leads to improved performance. In future work, we plan to explore advanced tokenization methods that explicitly target the memorization and generalization capabilities characterized in this study.

\bibliographystyle{assets/plainnat}
\bibliography{ref}

@inproceedings{rajput2023tiger,
  title={Recommender Systems with Generative Retrieval},
  author={Shashank Rajput and Nikhil Mehta and Anima Singh and Raghunandan Hulikal Keshavan and Trung Vu and Lukasz Heldt and Lichan Hong and Yi Tay and Vinh Q. Tran and Jonah Samost and Maciej Kula and Ed H. Chi and Maheswaran Sathiamoorthy},
  booktitle={{NeurIPS}},
  year={2023},
}

@article{he2025plum,
  author       = {Ruining He and
                  Lukasz Heldt and
                  Lichan Hong and
                  Raghunandan H. Keshavan and
                  Shifan Mao and
                  Nikhil Mehta and
                  Zhengyang Su and
                  Alicia Tsai and
                  Yueqi Wang and
                  Shao{-}Chuan Wang and
                  Xinyang Yi and
                  Lexi Baugher and
                  Baykal Cakici and
                  Ed H. Chi and
                  Cristos Goodrow and
                  Ningren Han and
                  He Ma and
                  R{\'{o}}mer Rosales and
                  Abby Van Soest and
                  Devansh Tandon and
                  Su{-}Lin Wu and
                  Weilong Yang and
                  Yilin Zheng},
  title        = {{PLUM:} Adapting Pre-trained Language Models for Industrial-scale
                  Generative Recommendations},
  journal      = {arXiv preprint arXiv:2510.07784},
  year         = {2025}
}

@article{deng2025onerec,
  author       = {Jiaxin Deng and
                  Shiyao Wang and
                  Kuo Cai and
                  Lejian Ren and
                  Qigen Hu and
                  Weifeng Ding and
                  Qiang Luo and
                  Guorui Zhou},
  title        = {OneRec: Unifying Retrieve and Rank with Generative Recommender and
                  Iterative Preference Alignment},
  journal      = {arXiv preprint arXiv:2502.18965},
  year         = {2025}
}

@inproceedings{ding2026specgr,
  title={Inductive Generative Recommendation via Retrieval-based Speculation},
  author={Ding, Yijie and Li, Jiacheng and McAuley, Julian and Hou, Yupeng},
  booktitle={{AAAI}},
  year={2026}
}

@article{yang2025liger,
  author       = {Liu Yang and
                  Fabian Paischer and
                  Kaveh Hassani and
                  Jiacheng Li and
                  Shuai Shao and
                  Zhang Gabriel Li and
                  Yun He and
                  Xue Feng and
                  Nima Noorshams and
                  Sem Park and
                  Bo Long and
                  Robert D. Nowak and
                  Xiaoli Gao and
                  Hamid Eghbalzadeh},
  title        = {Unifying Generative and Dense Retrieval for Sequential Recommendation},
  journal      = {Trans. Mach. Learn. Res.},
  volume       = {2025},
  year         = {2025}
}

@inproceedings{rendle2010fpmc,
  author    = {Steffen Rendle and
               Christoph Freudenthaler and
               Lars Schmidt{-}Thieme},
  title     = {Factorizing personalized Markov chains for next-basket recommendation},
  booktitle = {{WWW}},
  year      = {2010},
}

@inproceedings{zheng2024lcrec,
  title={Adapting Large Language Models by Integrating Collaborative Semantics for Recommendation},
  author={Bowen Zheng and Yupeng Hou and Hongyu Lu and Yu Chen and Wayne Xin Zhao and Ji-Rong Wen},
  booktitle={{ICDE}},
  year={2024}
}

@inproceedings{hou2025rpg,
  title={Generating Long Semantic IDs in Parallel for Recommendation},
  author={Yupeng Hou and Jiacheng Li and Ashley Shin and Jinsung Jeon and Abhishek Santhanam and Wei Shao and Kaveh Hassani and Ning Yao and Julian McAuley},
  booktitle={{KDD}},
  year={2025}
}

@inproceedings{hou2025actionpiece,
  title={{ActionPiece}: Contextually Tokenizing Action Sequences for Generative Recommendation},
  author={Yupeng Hou and Jianmo Ni and Zhankui He and Noveen Sachdeva and Wang-Cheng Kang and Ed H. Chi and Julian McAuley and Derek Zhiyuan Cheng},
  booktitle={{ICML}},
  year={2025}
}

@inproceedings{liu2025e2egrec,
  title={Generative recommender with end-to-end learnable item tokenization},
  author={Liu, Enze and Zheng, Bowen and Ling, Cheng and Hu, Lantao and Li, Han and Zhao, Wayne Xin},
  booktitle={{SIGIR}},
  pages={729--739},
  year={2025}
}

@inproceedings{hua2023p5cid,
  author       = {Wenyue Hua and
                  Shuyuan Xu and
                  Yingqiang Ge and
                  Yongfeng Zhang},
  title        = {How to Index Item IDs for Recommendation Foundation Models},
  booktitle    = {{SIGIR-AP}},
  year         = {2023}
}

@inproceedings{tay2022dsi,
  title={Transformer memory as a differentiable search index},
  author={Yi Tay and
  Vinh Tran and
  Mostafa Dehghani and
  Jianmo Ni and
  Dara Bahri and
  Harsh Mehta and
  Zhen Qin and
  Kai Hui and
  Zhe Zhao and
  Jai Prakash Gupta and
  Tal Schuster and
  William W. Cohen and
  Donald Metzler},
  booktitle={{NeurIPS}},
  year={2022}
}

@inproceedings{jin2024lmindexer,
  title={Language Models As Semantic Indexers},
  author={Bowen Jin and
  Hansi Zeng and
  Guoyin Wang and
  Xiusi Chen and
  Tianxin Wei and
  Ruirui Li and
  Zhengyang Wang and
  Zheng Li and
  Yang Li and
  Hanqing Lu and
  Suhang Wang and
  Jiawei Han and
  Xianfeng Tang},
  booktitle={{ICML}},
  year={2024}
}

@inproceedings{hidasi2016gru4rec,
  author    = {Bal{\'{a}}zs Hidasi and
               Alexandros Karatzoglou and
               Linas Baltrunas and
               Domonkos Tikk},
  title     = {Session-based Recommendations with Recurrent Neural Networks},
  booktitle = {{ICLR}},
  year      = {2016},
}

@inproceedings{kang2018sasrec,
  author    = {Wang{-}Cheng Kang and
               Julian J. McAuley},
  title     = {Self-Attentive Sequential Recommendation},
  booktitle = {{ICDM}},
  year      = {2018},
}

@inproceedings{sun2019bert4rec,
  author    = {Fei Sun and
               Jun Liu and
               Jian Wu and
               Changhua Pei and
               Xiao Lin and
               Wenwu Ou and
               Peng Jiang},
  title     = {BERT4Rec: Sequential Recommendation with Bidirectional Encoder Representations
               from Transformer},
  booktitle = {{CIKM}},
  year      = {2019},
}

@inproceedings{tang2018caser,
  author    = {Jiaxi Tang and
               Ke Wang},
  title     = {Personalized Top-N Sequential Recommendation via Convolutional Sequence
               Embedding},
  booktitle = {{WSDM}},
  year      = {2018}
}

@inproceedings{ma2019hgn,
  author       = {Chen Ma and
                  Peng Kang and
                  Xue Liu},
  title        = {Hierarchical Gating Networks for Sequential Recommendation},
  booktitle    = {{KDD}},
  year         = {2019}
}

@article{zhu2025beyond,
  title={Beyond Unimodal Boundaries: Generative Recommendation with Multimodal Semantics},
  author={Zhu, Jing and Ju, Mingxuan and Liu, Yozen and Koutra, Danai and Shah, Neil and Zhao, Tong},
  journal={arXiv preprint arXiv:2503.23333},
  year={2025}
}

@inproceedings{wang2025generative,
  title={Generative Next POI Recommendation with Semantic ID},
  author={Wang, Dongsheng and Huang, Yuxi and Gao, Shen and Wang, Yifan and Huang, Chengrui and Shang, Shuo},
  booktitle={{KDD}},
  year={2025}
}

@article{hou2024bridging,
  title={Bridging Language and Items for Retrieval and Recommendation},
  author={Hou, Yupeng and Li, Jiacheng and He, Zhankui and Yan, An and Chen, Xiusi and McAuley, Julian},
  journal={arXiv preprint arXiv:2403.03952},
  year={2024}
}

@inproceedings{www25-gen-rec-tutorial,
  author = {Hou, Yupeng and Zhang, An and Sheng, Leheng and Yang, Zhengyi and Wang, Xiang and Chua, Tat-Seng and McAuley, Julian},
  title = {Generative Recommendation Models: Progress and Directions},
  year = {2025},
  booktitle = {Companion Proceedings of the ACM Web Conference 2025},
}

@inproceedings{mcauley2015amazon,
  author       = {Julian J. McAuley and
                  Christopher Targett and
                  Qinfeng Shi and
                  Anton van den Hengel},
  title        = {Image-Based Recommendations on Styles and Substitutes},
  booktitle    = {{SIGIR}},
  year         = {2015}
}

@inproceedings{zhai2024hstu,
  title={Actions Speak Louder than Words: Trillion-Parameter Sequential Transducers for Generative Recommendations},
  author={Jiaqi Zhai and
                  Lucy Liao and
                  Xing Liu and
                  Yueming Wang and
                  Rui Li and
                  Xuan Cao and
                  Leon Gao and
                  Zhaojie Gong and
                  Fangda Gu and
                  Michael He and
                  Yinghai Lu and
                  Yu Shi},
  booktitle={{ICML}},
  year={2024}
}

@inproceedings{liu2024mbgen,
  title={Multi-Behavior Generative Recommendation},
  author={Zihan Liu and Yupeng Hou and Julian McAuley},
  booktitle={{CIKM}},
  year={2024}
}

@inproceedings{singh2024spmsid,
  title={Better Generalization with Semantic IDs: A Case Study in Ranking for Recommendations},
  author={Singh, Anima and Vu, Trung and Mehta, Nikhil and Keshavan, Raghunandan and Sathiamoorthy, Maheswaran and Zheng, Yilin and Hong, Lichan and Heldt, Lukasz and Wei, Li and Tandon, Devansh and Chi, Ed H. and Yi, Xinyang},
  booktitle={{RecSys}},
  year={2024}
}

@inproceedings{wang2024letter,
  title={Learnable Tokenizer for LLM-based Generative Recommendation},
  author={Wang, Wenjie and Bao, Honghui and Lin, Xinyu and Zhang, Jizhi and Li, Yongqi and Feng, Fuli and Ng, See-Kiong and Chua, Tat-Seng},
  booktitle={{CIKM}},
  year={2024}
}

@inproceedings{zhu2024cost,
  author       = {Jieming Zhu and
                  Mengqun Jin and
                  Qijiong Liu and
                  Zexuan Qiu and
                  Zhenhua Dong and
                  Xiu Li},
  title        = {CoST: Contrastive Quantization based Semantic Tokenization for Generative
                  Recommendation},
  booktitle    = {{RecSys}},
  year         = {2024}
}

@article{liu2024mmgrec,
  title={MMGRec: Multimodal Generative Recommendation with Transformer Model},
  author={Liu, Han and Wei, Yinwei and Song, Xuemeng and Guan, Weili and Li, Yuan-Fang and Nie, Liqiang},
  journal={arXiv preprint arXiv:2404.16555},
  year={2024}
}

@inproceedings{ju2025generative,
  title={Generative Recommendation with Semantic IDs: A Practitioner's Handbook},
  author={Ju, Clark Mingxuan and Collins, Liam and Neves, Leonardo and Kumar, Bhuvesh and Wang, Louis Yufeng and Zhao, Tong and Shah, Neil},
  booktitle={{CIKM}},
  pages={6420--6425},
  year={2025}
}

@article{zhang2024moc,
  title={Towards Scalable Semantic Representation for Recommendation},
  author={Taolin Zhang and
          Junwei Pan and
          Jinpeng Wang and
          Yaohua Zha and
          Tao Dai and
          Bin Chen and
          Ruisheng Luo and
          Xiaoxiang Deng and
          Yuan Wang and
          Ming Yue and
          Jie Jiang and
          Shu{-}Tao Xia},
  journal={arXiv preprint arXiv:2410.09560},
  year={2024}
}

@article{han2022images,
  title={What Images are More Memorable to Machines?},
  author={Han, Junlin and Zhan, Huangying and Hong, Jie and Fang, Pengfei and Li, Hongdong and Petersson, Lars and Reid, Ian},
  journal={arXiv preprint arXiv:2211.07625},
  year={2022}
}

@inproceedings{carlini2023quantifying,
  title={Quantifying Memorization Across Neural Language Models},
  author={Carlini, Nicholas and Ippolito, Daphne and Jagielski, Matthew and Lee, Katherine and Tramer, Florian and Zhang, Chiyuan},
  booktitle={{ICLR}},
  year={2023}
}

@article{yang2023resmem,
  title={ResMem: Learn what you can and memorize the rest},
  author={Yang, Zitong and Lukasik, Michal and Nagarajan, Vaishnavh and Li, Zonglin and Rawat, Ankit and Zaheer, Manzil and Menon, Aditya K and Kumar, Sanjiv},
  journal={Advances in Neural Information Processing Systems},
  volume={36},
  pages={60768--60790},
  year={2023}
}

@article{jiang2024investigating,
  title={Investigating data contamination for pre-training language models},
  author={Jiang, Minhao and Liu, Ken Ziyu and Zhong, Ming and Schaeffer, Rylan and Ouyang, Siru and Han, Jiawei and Koyejo, Sanmi},
  journal={arXiv preprint arXiv:2401.06059},
  year={2024}
}

@article{buchanan2025edge,
  title={On the edge of memorization in diffusion models},
  author={Buchanan, Sam and Pai, Druv and Ma, Yi and De Bortoli, Valentin},
  journal={arXiv preprint arXiv:2508.17689},
  year={2025}
}

@inproceedings{
wang2024generalization,
title={Generalization v.s. Memorization: Tracing Language Models{\textquoteright} Capabilities Back to Pretraining Data},
author={Xinyi Wang and Antonis Antoniades and Yanai Elazar and Alfonso Amayuelas and Alon Albalak and Kexun Zhang and William Yang Wang},
booktitle={{ICLR}},
year={2025},
}

@inproceedings{raunak2021curious,
  title={The curious case of hallucinations in neural machine translation},
  author={Raunak, Vikas and Menezes, Arul and Junczys-Dowmunt, Marcin},
  booktitle={{NAACL}},
  pages={1172--1183},
  year={2021}
}

@article{grosse2023studying,
  title={Studying large language model generalization with influence functions},
  author={Grosse, Roger and Bae, Juhan and Anil, Cem and Elhage, Nelson and Tamkin, Alex and Tajdini, Amirhossein and Steiner, Benoit and Li, Dustin and Durmus, Esin and Perez, Ethan and others},
  journal={arXiv preprint arXiv:2308.03296},
  year={2023}
}

@article{zhang2023counterfactual,
  title={Counterfactual memorization in neural language models},
  author={Zhang, Chiyuan and Ippolito, Daphne and Lee, Katherine and Jagielski, Matthew and Tram{\`e}r, Florian and Carlini, Nicholas},
  journal={Advances in Neural Information Processing Systems},
  volume={36},
  pages={39321--39362},
  year={2023}
}

@article{ghosh2025rethinking,
  title={Rethinking Memorization Measures and their Implications in Large Language Models},
  author={Ghosh, Bishwamittra and Das, Soumi and Wu, Qinyuan and Khan, Mohammad Aflah and Gummadi, Krishna P and Terzi, Evimaria and Garg, Deepak},
  journal={arXiv preprint arXiv:2507.14777},
  year={2025}
}

@inproceedings{xie2410memorization,
    title = "On Memorization of Large Language Models in Logical Reasoning",
    author = "Xie, Chulin  and
      Huang, Yangsibo  and
      Zhang, Chiyuan  and
      Yu, Da  and
      Chen, Xinyun  and
      Lin, Bill Yuchen  and
      Li, Bo  and
      Ghazi, Badih  and
      Kumar, Ravi",
    booktitle = "{IJCNLP}",
    year = "2025",
}

@article{djire2025memorization,
  title={Memorization or Interpolation? Detecting LLM Memorization through Input Perturbation Analysis},
  author={Djir{\'e}, Alb{\'e}rick Euraste and Kabor{\'e}, Abdoul Kader and Barr, Earl T and Klein, Jacques and Bissyand{\'e}, Tegawend{\'e} F},
  journal={arXiv preprint arXiv:2505.03019},
  year={2025}
}

@inproceedings{chu2025sft,
  title={Sft memorizes, rl generalizes: A comparative study of foundation model post-training},
  author={Chu, Tianzhe and Zhai, Yuexiang and Yang, Jihan and Tong, Shengbang and Xie, Saining and Schuurmans, Dale and Le, Quoc V and Levine, Sergey and Ma, Yi},
  booktitle={{ICML}},
  year={2025}
}

@article{barron2025too,
  title={Too Big to Think: Capacity, Memorization, and Generalization in Pre-Trained Transformers},
  author={Barron, Joshua and White, Devin},
  journal={arXiv preprint arXiv:2506.09099},
  year={2025}
}

@article{pruthi2020estimating,
  title={Estimating training data influence by tracing gradient descent},
  author={Pruthi, Garima and Liu, Frederick and Kale, Satyen and Sundararajan, Mukund},
  journal={Advances in Neural Information Processing Systems},
  volume={33},
  pages={19920--19930},
  year={2020}
}

@inproceedings{xia2024less,
  title={LESS: selecting influential data for targeted instruction tuning},
  author={Xia, Mengzhou and Malladi, Sadhika and Gururangan, Suchin and Arora, Sanjeev and Chen, Danqi},
  booktitle={{ICML}},
  pages={54104--54132},
  year={2024}
}

@article{vashurin2025benchmarking,
  title={Benchmarking uncertainty quantification methods for large language models with lm-polygraph},
  author={Vashurin, Roman and Fadeeva, Ekaterina and Vazhentsev, Artem and Rvanova, Lyudmila and Vasilev, Daniil and Tsvigun, Akim and Petrakov, Sergey and Xing, Rui and Sadallah, Abdelrahman and Grishchenkov, Kirill and others},
  journal={Transactions of the Association for Computational Linguistics},
  volume={13},
  pages={220--248},
  year={2025},
  publisher={MIT Press 255 Main Street, 9th Floor, Cambridge, Massachusetts 02142, USA~…}
}

@inproceedings{he2017neural,
  title={Neural collaborative filtering},
  author={He, Xiangnan and Liao, Lizi and Zhang, Hanwang and Nie, Liqiang and Hu, Xia and Chua, Tat-Seng},
  booktitle={{WWW}},
  pages={173--182},
  year={2017}
}

@article{ivison2025large,
  author       = {Hamish Ivison and
                  Muru Zhang and
                  Faeze Brahman and
                  Pang Wei Koh and
                  Pradeep Dasigi},
  title        = {Large-Scale Data Selection for Instruction Tuning},
  journal={arXiv preprint arXiv:2503.01807},
  year         = {2025}
}

@inproceedings{pezeshkpour2021empirical,
  author       = {Pouya Pezeshkpour and
                  Sarthak Jain and
                  Byron C. Wallace and
                  Sameer Singh},
  title        = {An Empirical Comparison of Instance Attribution Methods for {NLP}},
  booktitle    = {{NAACL}},
  pages        = {967--975},
  year         = {2021}
}

@inproceedings{
liu2401infini,
title={Infini-gram: Scaling Unbounded n-gram Language Models to a Trillion Tokens},
author={Jiacheng Liu and Sewon Min and Luke Zettlemoyer and Yejin Choi and Hannaneh Hajishirzi},
booktitle={{COLM}},
year={2024},
}

@article{zheng2025mtgrec,
  author       = {Bowen Zheng and
                  Enze Liu and
                  Zhongfu Chen and
                  Zhongrui Ma and
                  Yue Wang and
                  Wayne Xin Zhao and
                  Ji{-}Rong Wen},
  title        = {Pre-training Generative Recommender with Multi-Identifier Item Tokenization},
  journal      = {arXiv preprint arXiv:2504.04400},
  year         = {2025},
}

@article{zheng2025utgrec,
  author       = {Bowen Zheng and
                  Hongyu Lu and
                  Yu Chen and
                  Wayne Xin Zhao and
                  Ji{-}Rong Wen},
  title        = {Universal Item Tokenization for Transferable Generative Recommendation},
  journal      = {arXiv preprint arXiv:2504.04405},
  year         = {2025},
}

@article{zhong2025pctx,
  title={Pctx: Tokenizing Personalized Context for Generative Recommendation},
  author={Qiyong Zhong and Jiajie Su and Yunshan Ma and Julian McAuley and Yupeng Hou},
  journal={arXiv preprint arXiv:2510.21276},
  year={2025}
}

@article{zhang2025psid,
  title={Purely Semantic Indexing for {LLM}-based Generative Recommendation and Retrieval},
  author={Ruohan Zhang and Jiacheng Li and Julian McAuley and Yupeng Hou},
  journal={arXiv preprint arXiv:2509.16446},
  year={2025}
}

@article{liu2025gr_scaling,
  author       = {Jingzhe Liu and
                  Liam Collins and
                  Jiliang Tang and
                  Tong Zhao and
                  Neil Shah and
                  Clark Mingxuan Ju},
  title        = {Understanding Generative Recommendation with Semantic IDs from a Model-scaling
                  View},
  journal={arXiv preprint arXiv:2509.25522},
  year         = {2025},
}


\end{document}